# Tuning linear and nonlinear optical properties of wurtzite GaN by *c*-axial stress

Duanjun Cai[1] and Guang-Yu Guo [1, 2, 3]

[1]Department of Physics, National Taiwan University, Taipei 106, Taiwan

[2]Graduate Institute of Applied Physics, National Chengchi University, Taipei 116,

Taiwan

We study the linear and nonlinear optical properties of wurtzite GaN under *c*-axial stress field, using density functional theory calculations. The fully structural optimization at each c-axial strain was performed. The calculated dielectric functions show that tensile c-axial strain effectively improves the linear optical efficiency, especially for the band-edge transitions, and significantly increase the mobility of electrons in the conduction band. Second-order nonlinear optical susceptibilities show that the tensile c-axial strain will enhance the zero- and low-frequency nonlinear responses of GaN. The enhancement of the nonlinear optical property is explained by the reduction of the polarization of wurtzite GaN under tensile c-axial strains. Based on these findings, we propose a method for improving the electrical and optical properties of the crystal through imposing appropriate stress on the high symmetry crystalline directions.

[3] To whom corresponding should be addressed, e-mail: gyguo@phys.ntu.edu.tw

# 1. Introduction

The development of optoelectronic materials and high efficient devices today points to the need of further improvement of the optical and transport properties. Therefore, the attempts to grow high quality crystals with low defect density [1-3] or to design heterostructures for carrier confinement [4, 5] have been widely made. Theoretically, the optical or electric reaction can be understood as the process of electron transitions between different energy states [6]. The electronic energy band structure forms the basic frame for various electron movements in energy space. In particular, the important parameters such as the effective mass and the dielectric function can both be related to the gradient of the energy band, namely, the $\frac{\partial E}{\partial k_\alpha}$ or $\frac{\partial^2 E}{\partial k_\alpha^2}$, where $E$ is the electron energy and $k_\alpha$ is the wavevector in the reciprocal space [7]. This relation demonstrates that the shape of energy band is the key factor that determines the optoelectronic property of the material and therefore, band shape engineering could be an effective way to ameliorate the material properties. However, the problem is how to realize this engineering in practice and what rule should be followed.

In principle, the shape of energy band of a crystal is determined by the structural symmetry and electronic bonding state. For a material with known chemical composition and structural phase, actually, the ground state band structure has been determined. However, the stress field applied on the crystal will produce lattice distortion (strain) as well as bond bending and stretching. Thus, the point group symmetry along a crystallographic orientation will be changed, and the bonding energy and lattice constant will be modified. This would give rise to a deformation of the energy bands in the reciprocal space. Usually, along a high symmetry direction the curvature of the energy bands is smaller than that along a low symmetry one, as illustrated in figure 1. Therefore, the stress field along a high symmetry orientation is more likely to induce distinct band deformation, and hence changes the band gradient significantly. It is believed that a better optoelectronic property can be achieved by this mean.

Recently, GaN has attracted enormous interest because of their important application in short-wavelength and high-power optoelectronic devices [8-10]. The stable phase of GaN is wurtzite structure, which is an anisotropic structure defined by two lattice constants in perpendicular axes, *a* and *c*. The *c* orientation possesses a high symmetry group ($C_{6v}$) while *a* orientation has a lower one ($C_{2v}$) [11]. During the heteroepitaxial growth of GaN, misfit stress is inevitably generated due to the lattice mismatch and different thermal expansion between the epilayer and the substrate (e.g., sapphire is the widely used substrate). In general, they are in-plane (in *a* equivalent axes) misfit stresses while the out-of-plane (parallel to *c*-axis) stress has been greatly relaxed because of the existence of the free surface of GaN film. Therefore, the features and influences of the *c*-axial stress on GaN properties were rarely considered and studied.

In this work, we performed density functional theory (DFT) calculations on the electrical and optical properties of wurtzite GaN by introduction of c-axial stress field.

Firstly, the complete geometric optimization of GaN under a *c*-axial stress was carried out. Secondly, the absorptive spectra (i.e., the imaginary part of dielectric function) was calculated by using the Fermi's golden rule. Band-edge intensity enhancement and wavelength shift by *c*-strain was observed. Thirdly, transport property (i.e., the electron effective mass) was obtained from energy band calculation and it was found that tensile *c*-strain improves the electron mobility. Finally, computations of second-order nonlinear optical susceptibility were performed and the connection with polarization was discussed. Therefore, it was concluded that controlling the *c*-axial strain on GaN is an effective way to improve the electric and optical properties.

## 2. Theory and Calculation methods

Our DFT calculations for the bulk GaN were performed using highly accurate full-potential projector augmented wave (PAW) method [12], as implemented in the VASP package [13]. They are based on DFT with the local-density approximation (LDA) [14, 15]. The plane-wave cutoff energy was set to 550 eV and a (8×8×8) Monkhorst-Pack mesh was used to sample the Brillouin zone in geometric optimization and electronic structure calculations [16]. For the purpose of introducing stress, the equilibrium lattice constant $a_0$ for the wurtzite GaN have been determined to be 0.319 nm and $c_0/a_0 = 1.627$, which agree well with the reported results [17, 18].

The optical properties were calculated based on the dielectric function formula. The imaginary (absorptive) part of the dielectric function $\varepsilon(\omega)$ due to direct interband transitions is given by the Fermi's golden rule [19,20] (atomic units are used throughout in this paper), i.e.,

$$\varepsilon''_{aa}(\omega) = \frac{4\pi^2}{\Omega\omega^2} \sum_{i\in VB, j\in CB} \sum_{\mathbf{k}} w_\mathbf{k} \left|p_{ij}^a\right|^2 \delta(E_{\mathbf{k}j} - E_{\mathbf{k}i} - \omega),  \quad (1)$$

where $\Omega$ is the unit-cell volume, $\omega$ the photon energy, and $w_\mathbf{k}$ the weighting factor of corresponding $k$ point. Also, VB and CB denote the conduction and valence bands, respectively. The dipole transition matrix elements $p_{ij}^a = \langle \mathbf{k}j|\hat{p}_a|\mathbf{k}i\rangle$ were obtained from the self-consistent band structures within the PAW formalism. Here $|\mathbf{k}n\rangle$ is the $n$th Bloch state wave function with crystal momentum $\mathbf{k}$ and $a$ denotes the Cartesian components. The real part of the dielectric function is obtained from $\varepsilon''(\omega)$ by a Kramer-Kronig transformation

$$\varepsilon'(\omega) = 1 + \frac{2}{\pi}\mathbf{P}\int_0^\infty d\omega' \frac{\omega'\varepsilon''(\omega')}{\omega'^2 - \omega^2}. \quad (2)$$

Here **P** denotes the principal value of the integral. Given the complex dielectric function ($\varepsilon' + i\varepsilon''$), all other linear optical properties such as refractive index, reflectivity, and absorption spectrum can be calculated. Furthermore, the electron energy loss spectrum at the long-wavelength limit is $-\text{Im}[(\varepsilon' + i\varepsilon'')^{-1}]$ and the electric polarizability $\alpha$ is given by $\varepsilon'(\omega) = 1 + 4\pi\alpha(\omega)/\Omega$.

Following previous nonlinear optical calculations [19, 21], the imaginary part of the second-order optical susceptibility due to direct interband transitions is given by

$$\chi''^{(2)}_{abc}(-2\omega,\omega,\omega) = \chi''^{(2)}_{abc,VE}(-2\omega,\omega,\omega) + \chi''^{(2)}_{abc,VH}(-2\omega,\omega,\omega), \quad (3)$$

where the contributions due to the so-called virtual-electron (VE) process is

$$\chi''^{(2)}_{abc,VE} = -\frac{\pi}{2\Omega} \sum_{i \in VB} \sum_{j,l \in CB} \sum_{\mathbf{k}} w_{\mathbf{k}} \left\{ \frac{\text{Im}[p^a_{jl}\langle p^b_{li} p^c_{ij}\rangle]}{E^3_{jl}(E_{jl}+E_{ji})} \right.$$

$$\times \delta(E_{jl}-\omega) - \frac{\text{Im}[p^a_{ij}\langle p^b_{jl} p^c_{li}\rangle]}{E^3_{jl}(2E_{jl}-E_{ji})} \delta(E_{jl}-\omega) \quad (4)$$

$$\left. + \frac{16\,\text{Im}[p^a_{ij}\langle p^b_{jl} p^c_{li}\rangle]}{E^3_{ji}(2E_{jl}-E_{ji})} \delta(E_{ji}-2\omega) \right.$$

and that due to the virtual-hole (VH) process is

$$\chi''^{(2)}_{abc,VH} = \frac{\pi}{2\Omega} \sum_{i,l \in VB} \sum_{j \in CB} \sum_{\mathbf{k}} w_{\mathbf{k}} \left\{ \frac{\text{Im}[p^a_{li}\langle p^b_{ij} p^c_{jl}\rangle]}{E^3_{jl}(E_{jl}+E_{ji})} \right.$$

$$\times \delta(E_{jl}-\omega) - \frac{\text{Im}[p^a_{ij}\langle p^b_{jl} p^c_{li}\rangle]}{E^3_{jl}(2E_{jl}-E_{ji})} \delta(E_{jl}-\omega) \quad (5)$$

$$\left. + \frac{16\,\text{Im}[p^a_{ij}\langle p^b_{jl} p^c_{li}\rangle]}{E^3_{ji}(2E_{jl}-E_{ji})} \delta(E_{ji}-2\omega) \right.$$

Here $E_{ji} = E_{\mathbf{k}j} - E_{\mathbf{k}i}$ and $\langle p^b_{jl} p^c_{li}\rangle = \frac{1}{2}(p^b_{jl} p^c_{li} + p^b_{li} p^c_{jl})$. The real part of the second-order optical susceptibility is then obtained from $\chi''^{(2)}_{abc}$ by a Kramer-Kronig transformation

$$\chi'^{(2)}(-2\omega,\omega,\omega) = \frac{2}{\pi} \mathbf{P} \int_0^\infty d\omega' \frac{\omega' \chi''^{(2)}(2\omega',\omega',\omega')}{\omega'^2 - \omega^2}. \quad (6)$$

In the present calculations, the $\delta$ function in Eqs. (1) and (3) is approximated by a Gaussian function. The dense $k$-point sampling mesh of ($42 \times 42 \times 42$) was used. Furthermore, to ensure that $\varepsilon'$ and $\chi'^{(2)}$ calculated via Kramer-Kronig transformation [Eqs. (2) and (6)] are reliable, at least ten energy bands per atom are included in the present optical calculations.

The polarization calculation of GaN is based on the Berry phase theory of the electronic polarization of an insulating system [22, 23]. To obtain the electronic polarization, the dipole moment is calculated with the integration along three directions G1, G2 and G3 over the reciprocal space unit cell. Then the Born effective charges are obtained by summing the polarizations of these three directions. In order to get the accurate results, the number of k-point sampling along each reciprocal direction is set to 10. The electronic polarization of bulk GaN under different *c*-axial strain is calculated and the influence on optical properties is discussed.

**3. Structure and stress field**

GaN generally crystallizes in the most stable wurtzite structure, which has a hexagonal unit cell with two lattice parameters *a* and *c*. As shown in figure 2, this structure is composed of two hexagonal close-packed (*hcp*) sublattices, which are shifted with respect to each other along the three-fold *c* axis by the amount of *u*. *u* can be described by the Ga-N bond length or by *u/c* in fractional coordinates. These three parameters are needed to fully determine the wurtzite structure. Therefore, during the following investigations on the effects of *c*-axial strain, complete optimization of these parameters will be made.

Since the equilibrium lattice constants of $a_0$ and $c_0$ were optimized previously, *c*-axial stress was then imposed on the *c* orientation of bulk GaN by changing the *c* length. For the convenience of comparison, *c*-axial strain is used in the rest of this paper, which is defined as $\delta_c = \frac{c-c_0}{c_0} \times 100\%$. When the external stress is introduced on *c* axis, the corresponding distortion on *a* axis will occur as a result, which can also be defined as $\delta_a = \frac{a - a_0}{a_0} \times 100\%$. In general, the elastic strain on *a*-axis and *c*-axis can be correlated by the Poisson's ratio or the elastic constants. For a hexagonal crystal, it can be simply written as

$$\delta_a = -\nu \delta_c, \qquad (7)$$

$$\nu = \frac{C_{13}}{C_{11} + C_{12}}, \qquad (8)$$

where $\nu$ is the Poisson's ratio and $C_{11}$, $C_{12}$, and $C_{13}$ are the independent elastic constants. Although the above functions show the relation clearly, the Poisson's ratios of GaN reported in the literature are inconsistent with each other [24, 25]. Furthermore, during the lattice distortion by an external stress field, not only the parameter *c* and *a* lengths are changed but also the interval parameter *u* would be modified. For the above reasons and for the purpose of determining the optimized structure of GaN under a *c*-axial strain, both *a* and *u* constant are fully relaxed at each strain point. figure 3 shows $\delta_a$ as a function of $\delta_c$. One can see that, in fact, the ratio is not linear, as described the simple Poisson's ratio. It behaves like a slight quadratic curve of second order when the $\delta_c$ strain is large. Furthermore, the optimized parameter *u* shows a significant variation under different *c*-axial strains, as plotted in figure 4. It decreases in absolute bond length as the compressive c-axial strain increases, whereas increases in fraction coordinate *u/c*. The increase of fraction *u/c* shows that the bond length is rather sensitive to the *c*-axial strain and the ratio *u/c* reduction appears to be significant. This indicates that in the study of the effect by stress field, internal parameters like *u* also play an important role, so that the complete optimization is necessary for pertinent results.

## 4. Linear optical properties

Linear optical property in form of dielectric function ($\varepsilon' + i\varepsilon''$) was calculated for the bulk GaN under various *c*-axial strains by using Eqs. (1) and (2). First the $\varepsilon'$ and $\varepsilon''$ under stress-free condition was obtained, as shown in figure 5. Due to the shortcoming of LDA method, which usually gives a narrower band gap for semiconductors, the widely employed scissor operation with the energy shift of 1.4 eV was used to correct the LDA band gap. The strain-free dielectric function $\varepsilon$ agrees well with previous theoretical and experimental results [18, 26-28].

By imposing different *c*-axial strains, the systematic variation of the imaginary (absorptive) part of dielectric constant $\varepsilon'$ was calculated. It was found that the influence of the stress field on $\varepsilon''_c$ appeared more significant than that on $\varepsilon''_a$, as shown in the inset of figure 6. Therefore, we would like to focus on the $\varepsilon''_c$ spectra, especially on the most important band-edge absorptive peaks. Figure 6 shows the $\varepsilon''_c$ as a function of *c*-axial strain. One can clearly observe that the intensity of band-edge peak increases with increasing tensile *c*-axial strain whereas the intensity reduces with the compressive c-axial strain. In addition, the peak energy experiences a shift as the strain is imposed, which reflects the variation of energy band gap $E_g$ of GaN. To show the effect of *c*-axial strain on the band-edge peak, the intensity and corresponding band-gap width was plotted in figure 7. The band-edge intensity of $\varepsilon''_c$ increases almost linearly with the tensile c-axial strain. It is estimated that the increase of 1 % *c*-axial strain would induce about 4 % enhancement of band-edge light efficiency. On the other hand, the band gap becomes narrower with the expansion of *c*-axial strain, thereby giving rise to the red shift of the corresponding optical peak. These phenomena demonstrate that apart from varying the chemical composition of GaN-based ternary compound, the stress field in GaN can also significantly affect the optical properties, particularly for the case with stress field on the high symmetry axes. It can be another tunable parameter for fabrication of nitride-based optoelectronic devices.

**5. Dielectric constant and transport properties**

Figure 8 illustrates the dielectric constants $\varepsilon'_a$ and $\varepsilon'_c$ as a function of *c*-axial strain. $\varepsilon'_a$ exhibits the dielectric properties along the equivalent orientations within the (0001) plane whereas $\varepsilon'_c$ along the direction normal to the (0001) plane. Because the *a*-axial structure has been optimized and the strain relaxed, $\varepsilon'_a$ turns out to be less influenced by the *c*-strain, especially at the zero-frequency. For 4% strain, the $\varepsilon'_c$ in the zero-frequency limit is increased by 2.7% while the $\varepsilon'_a$ changes only within 0.2%. This means that the *a*-axial dielectric property is less susceptible to the c-axial treatments. In contrast, the $\varepsilon'_c$ constant is significantly affected. Such difference in the dielectric properties between two directions reflects the anisotropic feature of the hexagonal structure. Again, it is found that the tensile strain enhances the magnitude of $\varepsilon'_c$ while compressive one decreases it. In principle, $\varepsilon'$ is associated closely with the features such as refraction, oscillation, and polarization in solid. Therefore, from our results it can be understood that c-axial $\varepsilon'$ is more sensitive to structural circumstance such as stress field. This fact can be explained by the lack of mirror symmetry in *c* axis though it possesses a high symmetry group. Thus, the external strain will easily alter the dielectric properties, e.g., polarization, which will be further discussed later. On one hand, this seems to show the instability of *c*-axial dielectric, refractive, or polarization properties. On the other hand, the sensitivity provides us an opportunity to tune the properties on *c* axis.

In semiconductor materials, electrons in the conduction band and holes in the valence band can carry electrical current. The transport property can be described by the carrier mobility $\mu$ as a function of effective mass $m^*$: [29]

$$\mu = q\tau / m^*, \tag{9}$$

where $q$ is the charge of the carrier and $\tau$ the phenomenological scattering time. Usually, effective mass $m^*$ is defined in the unit of the electron mass $m_0$. In principle, the smaller carrier $m^*$ represents the larger carrier mobility. From the energy band calculations, $m^*$ for electrons in the conduction band ($m^*_e$) and holes in the valence band ($m^*_h$) are obtained as a function of *c*-axial strain, as shown in figure 9. It can be seen that $m^*_e$, for both *a* or *c* directions, decreases as the tensile strain increases and particularly, $m^*_{e,a}$ reduces more drastically. This indicates that under tensile *c*-strain, a higher electron mobility can be achieved. However, the situation for $m^*_h$ appears a little more complicated. For $m^*_{h,a}$, stress fields (both compressive one and tensile one) will reduce the hole effective mass. In the tensile c-strain region, the $m^*_{h,c}$ increases slightly so that the hole mobility parallel to *c* axis is decreased. While in the compressive region, surprisingly, the $m^*_{h,c}$ drops sharply down to nearly the same magnitude as that of the electron $m^*$, as the compressive *c*-axial strain increases. The reason of this special feature could be found in the heavy hole valence band near the Γ point and along the Γ-A direction, where a small shoulder appears and consequently gives rise to the rapid drop of $m^*_{h,c}$. The formation of the shoulder can be attributed to the band deformation by introduction of *c*-axial compressive stress. However, the detailed physics remains unclear.

## 6. Nonlinear optical properties and polarization effect

GaN has been observed as a potential nonlinear optical material, which exhibits significant second-order nonlinear responses [30, 31]. It was understood that polarization plays an important role in affecting the nonlinear susceptibility. Therefore, the stress induced variation in piezoelectric polarization would also be critical. Particularly, because the polarization constant of GaN exists along the *c* axis, *c*-axial stress naturally brings about direct impact on its nonlinear optical property. Here we perform the systematic calculations of real and imaginary parts of second-order nonlinear optical susceptibility $\chi^{(2)}(-2\omega, \omega, \omega)$ and electronic polarizations for different c-axial strains.

The susceptibility elements $\chi^{(2)}(-2\omega, \omega, \omega)$ of stress-free GaN system was calculated by using Eqs. (3) and (4). Wurtzite GaN has a noncentrosymmetric point group symmetry 6mm with a hexagonal primary cell. Thus, the three nonvanishing susceptibility tensor elements are of equivalent xxz = yyz = xzx = yzy, zxx = zyy, and zzz, as plotted in figure 10. Our results are consistent with previously reported results.[32] Obviously, the spectral peaks of $\chi^{(2)}_{zzz}(-2\omega,\omega,\omega)$ in the energy region just above the energy band gap appear more intense than those of $\chi^{(2)}_{xxz}$ and $\chi^{(2)}_{zxx}$. Therefore, we will focus mainly on the $\chi^{(2)}_{zzz}(-2\omega,\omega,\omega)$ spectrum below. figure 11 compares the absolute value of the imaginary part of $\chi^{(2)}_{zzz}$ and the imaginary part of dielectric function $\varepsilon''_c$, so that one can find out the origin of second-harmonic generation optical peaks. In the energy region between 2.0 and 7.0 eV [figure 11(a)], strikingly, almost all the peaks in $\chi^{(2)}_{zzz}$ can be one-to-one related to the peaks in $\varepsilon''_c(\omega/2)$ [figure 11(b)]. This indicates that the non-linear optical response in this energy region mostly comes from the double-photon resonance. In contrast, in the higher energy region above 7.0 eV, where no distinct $\varepsilon''_c(\omega/2)$ spectra can be assigned to, the $\chi''^{(2)}_{zzz}$ peaks can be roughly related to the $\varepsilon''_c(\omega)$ peaks. This indicates that they result from direct single-photon resonances. However, one can see that the correspondence between the $\chi''^{(2)}_{zzz}$ peaks and the $\varepsilon''_c(\omega)$ peaks in this high energy region appears not as good as that in the lower energy region. This suggests the complexity for the high energy peaks in $\chi''^{(2)}_{zzz}$ where the tails from $\varepsilon''_c(\omega/2)$ or other low energy photo-excitations might also be involved in the resonances.

Figure 12 displays the imaginary part of second-order optical susceptibility $\chi''^{(2)}_{zzz}$ in the low energy region as a function of *c*-axial strain. The variation of $\chi''^{(2)}_{zzz}$ due to *c*-strain appears rather pronounced. The band-edge peak at around 3.4 eV is enhanced by the tensile strain whereas reduced rapidly with the compressive strain. The tensile-strain induced enhancement has been found to be about 5.8% per 1% strain, which is more significant than that for the linear optical enhancement discussed in Sec. IV. This may be due to the generation of non-linear optical peaks by the process of double-photon resonance, by which the enhancement might be double. For the same reason, the intensity reduction by the compressive strain is also double. As a result, the peak at about 4.4 eV nearly disappears when the c-axial compressive strain becomes larger than 4%. The values below the energy band gap represent the zero- and low-frequency nonlinear response, which can be seen more clearly in the real part of second-order susceptibility ($\chi'^{(2)}_{zzz}$), as shown in figure 13. For $\delta_c = 0$, the value of zero-frequency limit $\chi'^{(2)}_{zzz}|_{\omega \to 0}$ is listed in Table 1 for comparison with previous experimental [33-35] and theoretical [32, 36, 37] results. Although the measured susceptibility is not conclusive because of the large experimental uncertainties for wurtzite GaN, the value of $\chi'^{(2)}_{zzz}|_{\omega \to 0}$ in this work agrees rather well with the reported results. Under c-axial strains, one can find that the enhancement of $\chi'^{(2)}_{zzz}|_{\omega \to 0}$ by tensile strain is significant. For 4% c-strain, the $\chi'^{(2)}_{zzz}|_{\omega \to 0}$ has been increased up to the amount of about $3 \times 10^{-8}$ esu. This not only explains the experimental uncertainties, where the stress situation appears complicated in different samples, but also provides a good solution for the enhancement of the nonlinear optical properties of GaN by controlling the stress field rather than the complicated design of the multilayers such as AlGaN/GaN heterostructures [38].

In general, the second-order nonlinear optical susceptibility is related to the polarization property. Therefore, the calculated electronic polarization of bulk GaN under c-axial strains is shown in figure 14. One can see that the polarization constant reduces with the tensile *c*-axial strain whereas increases with the compressive one. This is mainly due to the variation of the piezoelectric contribution to total polarizations. The total polarization for a piezoelectric material is the summation of the components of spontaneous ($P_{sp}$) and piezoelectric ($P_{pz}$) polarization. Wurtzite GaN naturally yields spontaneous polarization $P_{sp}$ as a result of its lack of inversion symmetry, the direction of which is parallel to [000$\bar{1}$].[36] Under tensile c-axial strains, $P_{pz}$ of wurtzite nitride is orientated in the [0001] direction and the magnitude is proportional to the strain. Therefore, the tensile *c*-axial strain-induced increase of $P_{pz}$ partially cancels the $P_{sp}$ and hence the total polarization is reduced, as shown in figure 14. In contrast, the compressive *c*-strain gives rise to an increase of total polarization. By comparison with the above results of $\chi_{zzz}^{(2)}$, we can find that the reduction of total polarization correlates with the enhancement of nonlinear optical properties while the increase of polarization reduces them.

## 7. Conclusions

In summary, we have performed DFT calculations on the optical and transportation properties of wurtzite GaN under *c*-axial stress field. The structural relaxation with c-axial strains showed that the full relaxation including *a*-axial strain and internal parameter *u* is important for achieving the optimized structure with a fixed *c*-axial strain. The dispersive (real) and the absorptive (imaginary) part of the dielectric function were calculated by using the Fermi golden rule. It was found that a tensile c-axial strain can effectively enhance the optical efficiency, for the band-edge peak, by about 4% with the increase of 1% *c*-axial strain. Furthermore, the blue shift of the main band-edge transitions is also caused by this tensile strain. It was also observed under tensile *c*-axial strain that the transport properties are improved with a better dielectric constant $\varepsilon'$ and smaller electron effective mass $m^*_e$. Finally, the second-order nonlinear optical susceptibilities were calculated. The peaks of $\chi'''^{(2)}_{zzz}$ in the low energy region are found to be correlated with the double-photon resonances whereas the peaks in the high energy region are associated with the single-photon resonances. Tensile c-axial strain was found to enhance the nonlinear optical intensity of both zero-frequency and low-frequency. The improvement of the band-edge peaks reaches about 5.8% for the increase of 1% *c*-strain. This is explained in terms of the calculated electronic polarization in GaN that the tensile c-axial strain will reduce the total polarization and thus the linear and nonlinear optical properties are enhanced. Based on these findings, we conclude that the stress field imposed on the high symmetry axes such as the *c*-axis of wurtzite GaN, can significantly change the optical and electrical properties.

**Acknowledgements**

The authors acknowledge supports from National Science Council, TSMC under TSMC JDP # NTU-0806 and NCTS of Taiwan. They also thank NCHC of Taiwan and NTU Computer and Information Networking Center for providing CPU time.

Table 1. Static seconde-order nonlinear optical susceptibility $\chi_{zzz}^{(2)}$ of GaN. $d$ represents the thickness of GaN film used in the experiments.

|       |           | $d$ (μm)    | $\left|\chi_{zzz}^{(2)}\right|$ (pm/V) |
|-------|-----------|-------------|----------------------------------------|
| Expt. | Ref. [33] | 0.3         | 9.64±0.6                               |
|       | Ref. [31] | 0.7 ~ 5.31  | 10.7                                   |
|       | Ref. [32] | 2.134       | 7.4                                    |
|       |           | 4.40        | 10.4                                   |
|       |           | 160.14      | 8.0                                    |
|       |           | 229.67      | 9.2                                    |
| Calt. | Ref. [34] | -           | 7.0                                    |
|       | Ref. [30] | -           | 10.42                                  |
|       | This work | -           | 8.38                                   |

**FIGURE CAPTIONS**

**Figure 1.** Schematic diagram of band structure. Along a low symmetry direction the band is steeper whereas along a high symmetry direction it is flatter.

**Figure 2.** Unit cell of wurtzite structure of GaN. Parameters *a*, *c* and *u* are shown.

**Figure 3.** Calculated strain on the *a* axis ($\delta_a$) as a function of the strain on the *c* axis ($\delta_c$).

**Figure 4.** Internal parameter *u* as a function of c-axial strain $\delta_c$. Left coordinate in absolute bond length (solid square) and right one in fraction (hollow triangle).

**Figure 5.** (color online) Real (dash and dash dot lines) and imaginary (solid and dotted lines) parts of dielectric function of GaN without stress.

**Figure 6.** Imaginary (absorptive) part of dielectric function parallel to *c* axis ($\varepsilon''_c$) as a function of *c*-axial strain $\delta_c$. The inset is the *a*-axial $\varepsilon''_a$ under different c-axial strains $\delta_c$. The dotted lines correspond to the curve under zero strain condition.

**Figure 7.** Band gap $E_g$ (hollow square) and intensity of $\varepsilon''_c$ for the band-edge transition (solid triangle), respectively, as a function of $c$-axial strain $\delta_c$.

**Figure 8.** Real (dispersive) part of dielectric function parallel to $a$ ($\varepsilon'_c$) and $c$ axis ($\varepsilon'_c$), (a) and (b), respectively, as a function of $c$-axial strain $\delta_c$. The dash dot and dotted lines correspond to the curve under zero strain condition.

**Figure 9.** (color online) Effective mass $m^*$ of electrons in the conduction band ($m^*_e$) (solid lines) and holes in the valence band ($m^*_h$) (dotted lines) as a function of c-axial strain $\delta_c$.

**Figure 10.** Imaginary part of second-order nonlinear optical susceptibility $\chi''^{(2)}(-2\omega,\omega,\omega)$. The nonvanishing tensor elements are $\chi''^{(2)}_{xzx}$, $\chi''^{(2)}_{zxx}$, and $\chi''^{(2)}_{zzz}$.

**Figure 11.** (a) Absolute value of imaginary part of second-order nonlinear optical susceptibility $|\chi''^{(2)}_{zzz}(-2\omega,\omega,\omega)|$, (b) absorptive part of dielectric function $\varepsilon''_c(\omega)$ (solid line) and $\varepsilon''_c(\omega/2)$ (dotted line). The origin and correspondence of peaks of $|\chi''^{(2)}_{zzz}(-2\omega,\omega,\omega)|$ from $\varepsilon''_c$ are identified.

**Figure 12.** Imaginary part of second-order nonlinear optical susceptibility $\chi''^{(2)}_{zzz}$ as a function of $c$-axial strain $\delta_c$. The dotted line corresponds to the curve under zero strain condition.

**Figure 13.** Real part of second-order nonlinear optical susceptibility $\chi'^{(2)}_{zzz}$ as a function of c-axial strain $\delta_c$. The dotted line corresponds to the curve under zero strain condition.

**Figure 14.** Polarization $P$ of GaN as a function of c-axial strain $\delta_c$.

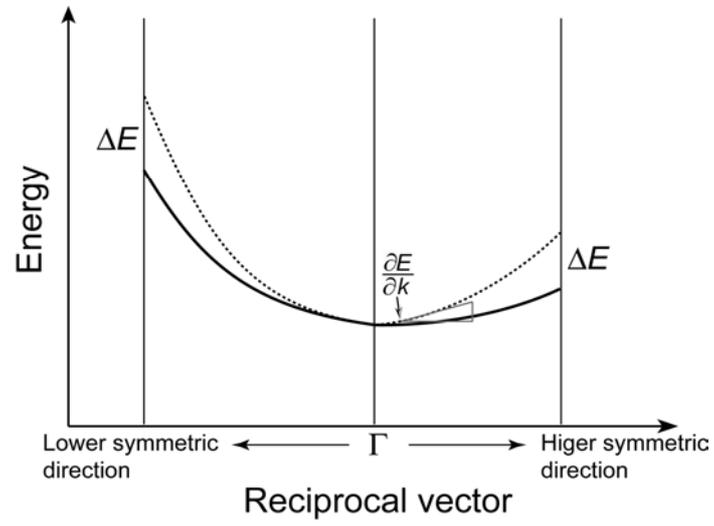

Figure 1. D. Cai et al

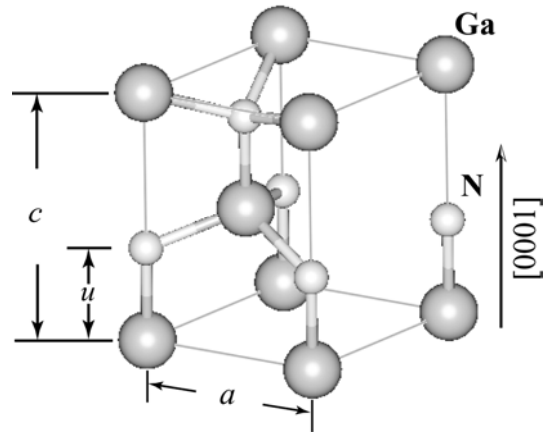

Figure 2. D. Cai et al

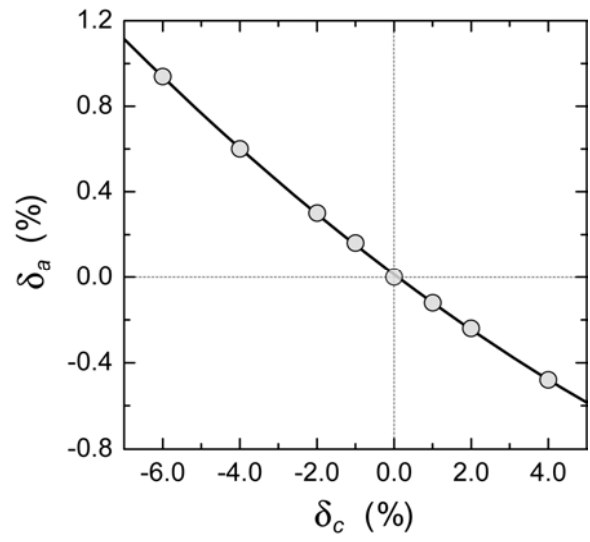

Figure 3. D. Cai et al

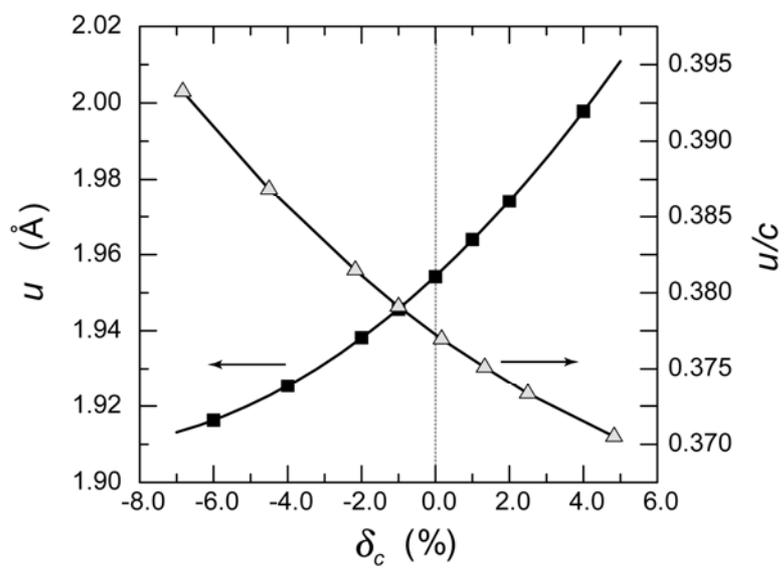

Figure 4. D. Cai et al

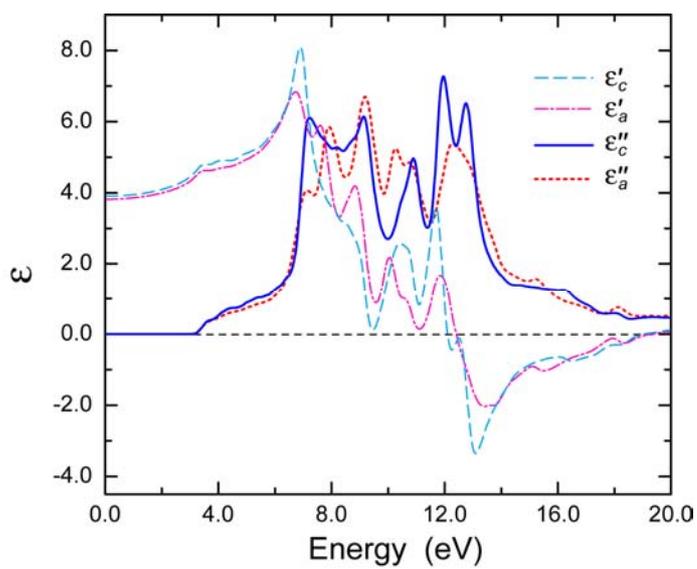

Figure 5. D. Cai et al

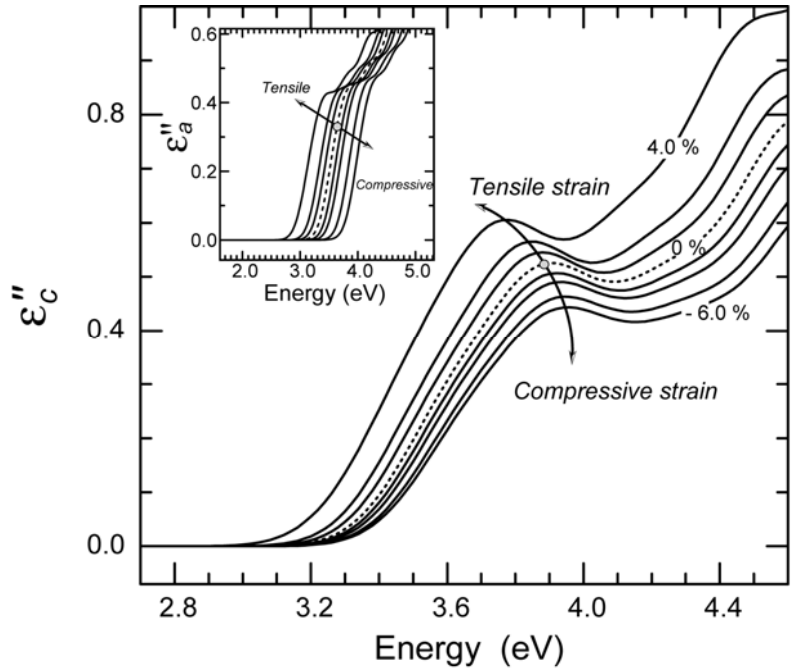

Figure 6. D. Cai et al

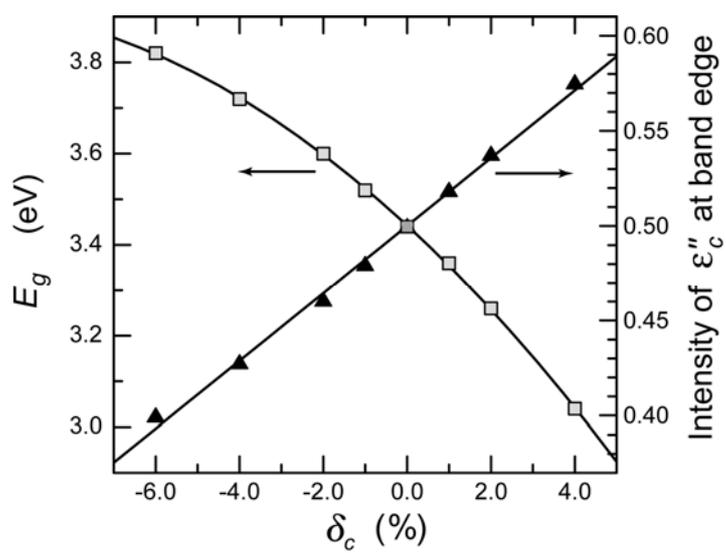

Figure 7. D. Cai et al

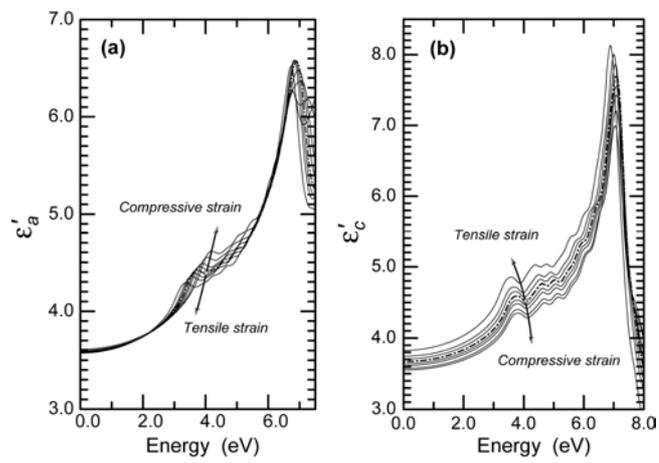

Figure 8. D. Cai et al

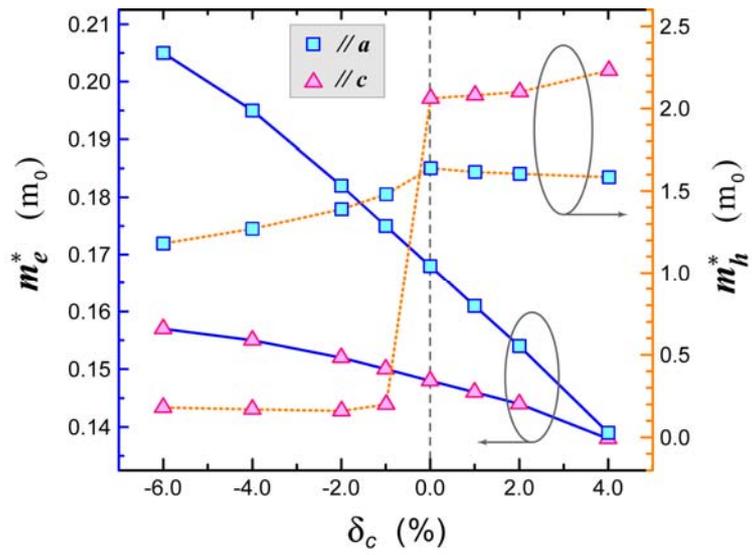

Figure 9. D. Cai et al

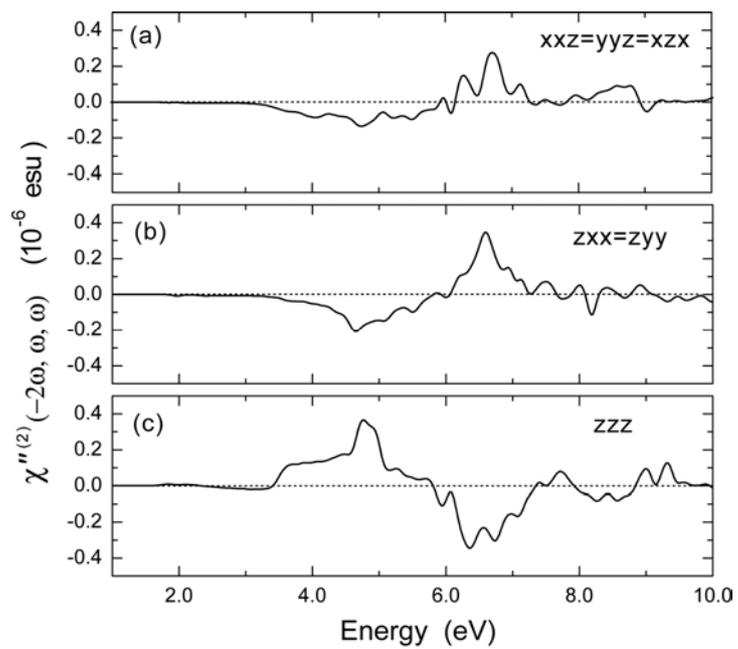

Figure 10. D. Cai et al

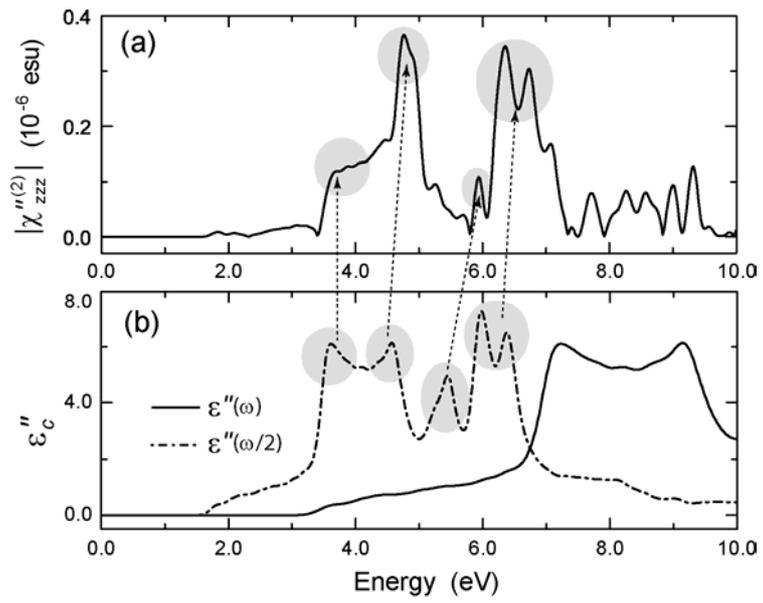

Figure 11. D. Cai et al

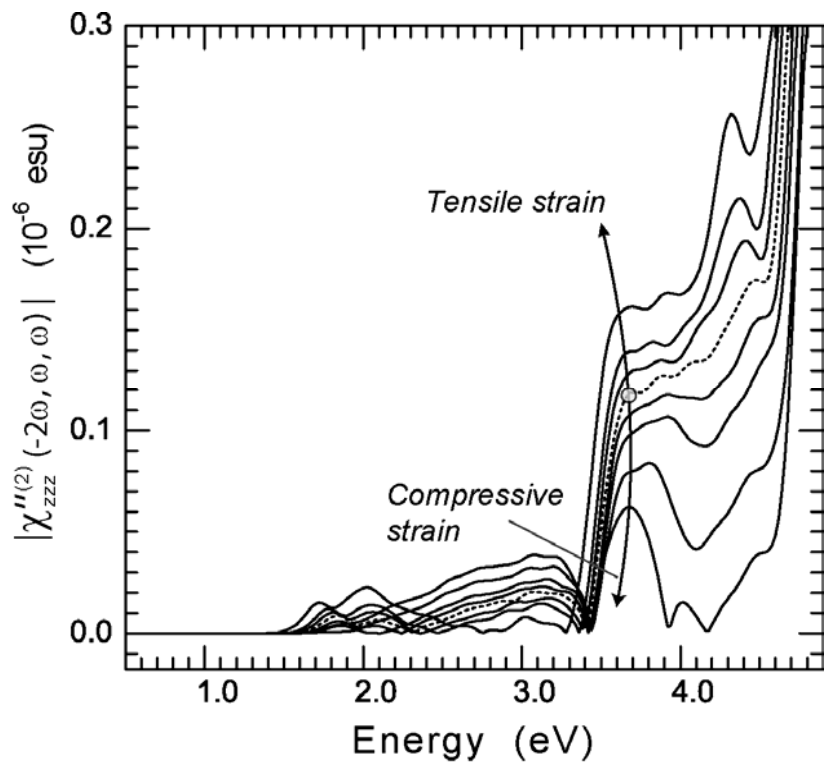

Figure 12. D. Cai et al

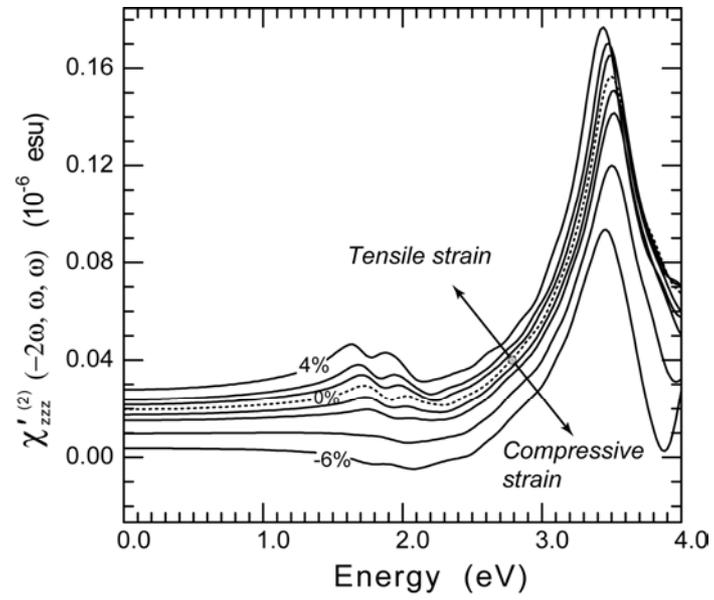

Figure 13. D. Cai et al

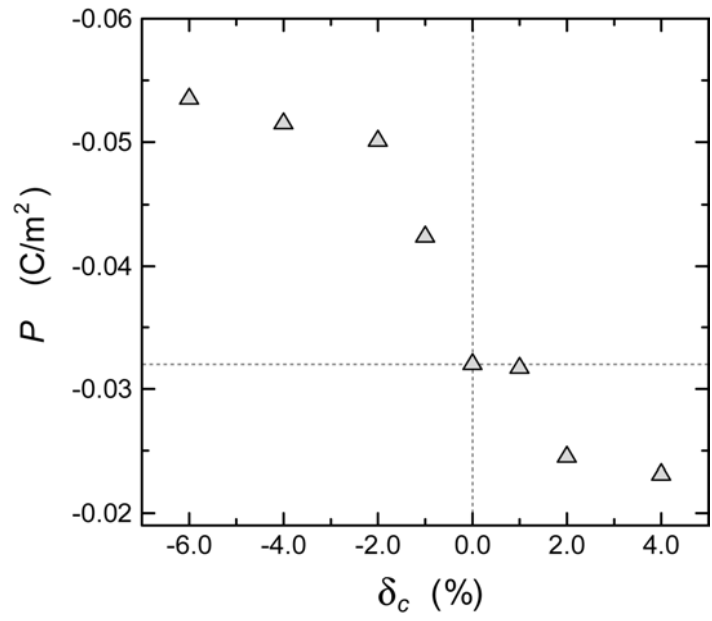

Figure 14. D. Cai et al